# Dimensionality-controlled evolution of charge-transfer energy in digital nickelates superlattices

*Xiangle Lu, Jishan Liu\*, Nian Zhang, Binping Xie, Shuai Yang, Wanling Liu, Zhicheng Jiang, Zhe Huang, Yichen Yang, Jin Miao, Wei Li, Soohyun Cho, Zhengtai Liu, Zhonghao Liu, and Dawei Shen\**

X. Lu, Prof. J. Liu, Dr. N Zhang, S. Yang, W. Liu, Z. Jiang, Z. Huang, Y. Yang, Dr. S. Cho, Dr. Z. Liu, Prof. Z. Liu, Prof. D. Shen
State Key Laboratory of Functional Materials for Informatics, Shanghai Institute of Microsystem and Information Technology, Chinese Academy of Sciences, Shanghai 200050, China
Center of Materials Science and Optoelectronics Engineering, University of Chinese Academy of Sciences, Beijing 100049, China

E-mail: jishanliu@mail.sim.ac.cn
dwshen@mail.sim.ac.cn

Dr. B. Xie
Feimion Instruments (shanghai) Company, Limited, Shanghai 201906, China

J. Miao, Dr. W. Li
State Key Laboratory of Surface Physics, Department of Physics, Fudan University, Shanghai 200433, China

Fundamental understanding and control of the electronic structure evolution in rare-earth nickelates is a fascinating and meaningful issue, as well as being helpful to understand the mechanism of recently discovered superconductivity. Here we systematically study the dimensionality effect on the ground electronic state in high-quality $(NdNiO_3)_m/(SrTiO_3)_1$ superlattices through transport and soft x-ray absorption spectroscopy. The metal-to-insulator transition temperature decreases with the thickness of the $NdNiO_3$ slab decreasing from bulk to 7 unit cells, then increases gradually as m further reduces to 1 unit cell. Spectral evidence demonstrates that the stabilization of insulating phase can be attributed to the increase of the charge-transfer energy between O $2p$ and Ni $3d$ bands. The prominent multiplet feature on the Ni $L_3$ edge develops with the decrease of $NdNiO_3$ slab thickness, suggesting the strengthening of the charge disproportionate state under the dimensional confinement. Our work provides convincing evidence that dimensionality is an effective knob to modulate the charge-transfer energy and thus the collective ground state in nickelates.

**1. Introduction**

Rare-earth nickelates (RENiO$_3$) are fascinating materials which exhibit a wide range of exotic phenomena originating from the interplay among charge, spin, orbital and lattice degrees of freedom. Among their many exotic properties, the metal-to-insulator transition (MIT) is perhaps the most appealing one. [1-4] Particularly, the MIT temperature ($T_{MIT}$) can be continuously tuned by changing rare-earth ions, [5-7] external strain, [8-13] charge transfer at heterostructure interface, [14-16] length scales of interfacial coupling. [3] Moreover, current advances in thin film growth technologies make it possible to obtain atomically precise digital oxide heterostructures and superlattices (SLs), which opens the new window for investigating strongly correlated electron systems in confined geometries. [17-19] In this context, artificial nickelate SLs would pave a promising way to explore and manipulate the metal-to-insulator transition by interface and dimensionality. While, the physical mechanism by which dimensionality modulates metal-to-insulator transition is very complex, the dimensionality could induce readjustment of octahedral rotations, electronic correlations, etc. A key ingredient in discovering the nature of metallic and insulating states is exploring the electronic states. Thus, it is vital to understand the dimensionality effect on the ground electronic state through systematic investigation of the MIT in nickelate SLs. The study of these mechanisms can significantly facilitate the design and application of devices based on nickelates.

Furthermore, the recent discovery of superconductivity in Nd$_{0.8}$Sr$_{0.2}$NiO$_2$ epitaxially grown on SrTiO$_3$ (STO) has greatly simulated the study of superconductivity and correlations in transition metal nickelates. [20-25] To date, the mechanism of nickel-based superconductors is still exclusive, although the consensus that the interface between Nd$_{0.8}$Sr$_{0.2}$NiO$_2$ and STO and dimensionality of Nd$_{0.8}$Sr$_{0.2}$NiO$_2$ should be intimately linked with the superconductivity has been achieved since the bulk doped NdNiO$_3$ (NNO) is not superconducting. [26, 27] Consequently, this fact greatly promoted the progress on the synthesis and characterization of ultra-thin films and heterostructures of rare-earth nickelates.

In this work, we report the epitaxial growth of a series of high quality [(NdNiO$_3$)$_m$/(SrTiO$_3$)$_1$] (m =1, 2, 3, 4, 5, 7, 9) SLs using reactive molecular beam epitaxy (MBE). we then systematically investigated the electronic structure evolution with the thickness of NNO slab decrease in superlattice structure through transport and soft x-ray absorption spectroscopy (XAS). Here, the dimensionality can be effectively tuned through changing m in one unit cell of SLs. The $T_{MIT}$ decreases firstly when the thickness of the NNO slab is reduced from bulk to 7 unit cells, and then increases gradually until out of the measurement range up to 400 K from m = 7 to 1 unit cell. Soft XAS results reveal that the charge-transfer energy, which is the energy cost for transferring an electron/hole from O 2$p$ to Ni 3$d$ bands, increases significantly with the thickness of NNO reducing, leading to the stabilization of the insulating phase in thinner NNO slabs. Meanwhile, prominent multiplet features on the Ni $L_3$ edge developed with the decrease of NdNiO$_3$ slab thickness, which suggests the charge disproportionate states are strengthened. Our work provides compelling evidence that the dimensionality is an effective way to modulate the charge-transfer energy between Ni 3$d$ and O 2$p$, and plays an indispensable role in understanding the physical mechanism of dimensionality effect on the MIT of perovskite nickelates.

**2. Results and discussion**

**Figure 1**(a-c) schematically show crystal structure of NNO/STO SLs for m = 1, 2 and 3. Their corresponding reflection high-energy electron diffraction (RHEED) [100] diffraction patterns are illustrated in Figure 1(d-f), respectively. These RHEED patterns exhibit sharp, bright and unmodulated streaks and no additional spots can be observed, indicating their atomically flat surfaces and high-quality crystallinity. Figure 1(d) presents a typical oscillation

of RHEED intensity integrated over the [00] diffraction rod for the m = 5 SL. Each growth cycle is highlighted in bright green and dark yellow, which represent the growth periods for one layer of STO and 5 layers of NNO, respectively. The regular RHEED oscillation (more details shown in the left lower enlarged inset) unambiguously demonstrates the layer-by-layer growth mode and atomic-scale flatness for each layer during the whole epitaxial growth. The right upper inset of Figure 1(g) shows the atomic force microscope (AFM) morphology for the m = 5 SL. The average surface roughness is only 0.13 nm and the terrace-step surface inherited from the substrate is clearly visible after the deposition, which further confirms the atomically flat surface of our SLs and layer-by-layer growth mode.

To further characterize the phase purity and crystallinity of these SLs, high-resolution X-ray diffraction (XRD) measurements were carried out. **Figure 2** shows the out-of-plane $\theta$-$2\theta$ scans around the STO (002) peak for m = 1, 2, 3, 4 and 5 SLs, respectively. Besides the prominent fundamental peaks, clear Bragg peaks corresponding to the periodicity of (m+1) a (the lattice constant a = 3.905 Å) demonstrate the phase-pure and epitaxial (001) oriented SLs. Furthermore, Laue oscillations around the SLs main diffraction peaks are clearly visible which suggest lateral homogeneity and well-defined interface between the epitaxial films and the substrate. The inset of Figure 2 illustrates the rocking curve around the fundamental peak for m = 3 SL, which is a universally agreed-on means to characterize the crystalline quality of epitaxial films. The full width at half maximums (FWHM) is only about 0.062° through Gaussian fitting, revealing the superior crystalline perfection of this SL. Note that all the SLs in this work exhibit similar characteristic RHEED patterns, AFM morphologies and FWHMs of rocking curves, as shown in **Figure S1-S3** in our supporting information, demonstrating the well-controlled thickness, smooth surface, sharp interface and high-quality crystallinity of our samples. Moreover, all the films grown on STO substrates are fully strained without relaxation, as supported in **Figure S4** and **S5.**

**Figure 3**(a) displays the in-plane resistivity as a function of temperature ranged from 10 K to 300 K. A systemic change of the MIT was obtained across the whole SLs series. The $T_{MIT}$ decreases firstly when the thickness of NNO slab is reduced from bulk to m = 7, and then increases gradually until out of the measurement range up to 400K from m = 7 to 1, as shown in Figure 3(b), in which the $T_{MIT}$ of different SLs were obtained through derivation $d\rho/dT = 0$ form Figure 3(a). Note that the hysteresis in resistivity is markedly weakened upon reducing m (**Figure S7**), which could be ascribe to the suppression of the structural change under the dimensional confinement. Here, the decrease of $T_{MIT}$ with the thickness of NdNiO$_3$ slab reducing from bulk to 7 unit cells can be attributed to the tensile stain effect. The tensile strain could straighten the Ni-O-Ni bond angle, and the orbital overlap is thus enhanced between Ni 3*d* states and O 2*p* states, which results in the increase of the bandwidth and thus the decrease of $T_{MIT}$.[28-30] While, with the thickness decreasing further, we speculate that dimensionality or quantum confinement effect would dominate the ground electronic state, which leads to the carriers localization behavior for ultra-thin NNO slabs.

To understand the resistivity evolution with m, we conducted a fitting analysis on these resistivity curves using three transport models of variable range hopping, small polaron hopping and activated behavior (also known as nearest neighbors hopping), as shown in Figure 3 (c), (d) and (e), respectively. The resistivity for m = 2 SL was found best fitted with the activated behavior model at high temperatures above 118 K. The activated behavior is an activation process in semiconducting systems, which follows $\rho(T) \propto e^{E_g/k_B T}$, where $E_g$ is the thermal activation energy of hopping electrons, and the $k_B$ is Boltzmann constant. The fitting result indicates a gap opened with m reducing into a limit value.[31, 32]

Nextly, soft XAS measurements was performed to gain microscopic insight into the effect of dimensionality and quantum confinement on the nature of MIT. **Figure 4**(a) shows the normalized O K-edge XAS pre-edge features for these SLs. These prepeaks are located at around 529 eV, corresponding to the transition from the O 1*s* core to the lowest unoccupied hybridized O 2p-Ni 3d band. We observed that these prepeaks exhibit a remarkable energy shift towards lower photon energy with m reducing from m = 7 to 1, as shown in Figure 4(a) and (b). The positions of prepeak were determined by the charge-transfer energy ($\Delta$), which is the energy cost for transferring an electron/hole from the O 2p band to the Ni 3d band. [33, 34] The remarkable shift of oxygen pre-edge highlights an increase of $\Delta$ when the NNO slab reducing. The increment of charge-transfer energy could be further revealed by the Ni L edge spectroscopy. The normalized spectra of Ni $L_3$ edge are illustrated in Figure 4(c). The Ni $L_3$ edge displays a splitting into two primary peaks, and the splitting becomes more and more pronounced with m reducing. Such peak splitting has been associated with the charge-transfer energy separating the O 2p and Ni 3d states near the Fermi level in previous literature. [32, 35, 36] To quantify the observed multiplet splitting energy, we fit the Ni $L_3$ spectra with the sum of two peaks for all SLs. The Ni $L_3$ splitting energy becomes enlarged as m reducing, indicating the increase of charge-transfer energy, as demonstrated in Figure 4(e). This finding is in good agreement with the observed shift of O K-edge prepeaks. Note that no charge transfer happened at the STO/NNO interface for the Ti cation strongly prefers the +4 oxidation state regardless of the thickness of NNO slab, as shown in Figure S8.

Thus, we can elucidate the increase of $T_{MIT}$ observed in transport behaviors. As well known, the extent of metal 3d-oxygen 2p hybridization scales with the ratio of the transfer integral to charge-transfer energy. [14] For nickelates, the Ni 3d-O 2p hybridization is primarily controlled by charge-transfer energy, while the transfer integral has a weaker influence. [37] Therefore, the increase of $\Delta$ with m decreasing would significantly weaken the Ni 3d-O 2p hybridization, resulting in the stabilization of the insulating phase. In addition, the observed multiplet feature on the Ni L edges in our case is closely reminiscent of absorption spectra of dimensional-confined heterostructured LaNiO3 slabs and other members of RENiO3 series with smaller RE ions in the insulating phase, [38-41] which have been identified as a signature of localized carriers in insulating charge-ordered nickelates. This inspires us to conjecture that the carriers gradually tend to be localized caused by charge ordering as the NNO slab reducing.

While, it is worth noting that nickelates are well recognized as negative charge transfer materials because of strong hybridization between the oxygen 2$p$ and nickel 3$d$ bands. [5, 42, 43] The ground electronic state of Ni$^{3+}$ is more likely as $d^8\underline{L}$. In that case, the electronic disproportionation (bond and charge) can be described as $d^8\underline{L} + d^8\underline{L} \rightarrow d^8 + d^8\underline{L}^2$.[43] This picture has been confirmed by both theories and experiments, [42, 44-46] providing an understanding of the metal-to-insulator transition as a site-selective Mott transition, in which the charge disproportionation takes place on the oxygen sites.[47] This model differs from the traditional charge-disproportionation ones wherein the Ni 3$d^7$ moves towards an alternation of Ni 3$d^{7-\delta}$ and 3$d^{7+\delta}$ sites in the insulating phase. In this scenario, the nickelates undergo an effective charge diproportionation achieved without any significant movement of charge between neighboring NiO$_6$ octahedra, and the insulating phase results from a partial volume collapse of oxygen octahedra with two ligand holes around their central Ni, while the other octahedra expand accordingly with little net effect on the total volume. [45, 46]

## 3. Conclusion

High quality (NdNiO$_3$)$_m$/(SrTiO$_3$)$_1$ superlattices were grown to study the dimensionality effect on the ground electronic state of NdNiO$_3$. The metal-to-insulator transition temperature decrease firstly, then increase gradually as the thickness of NNO slab decrease from bulk to 1 unit cell. Microscopically, the charge transfer energy between O 2$p$ and Ni 3$d$ was confirmed to increase as the thickness of the NNO slab reduced from 7 to 1 unit cell, which lead to the metal-insulator transition temperature gradually increasing until the superlattice becomes an insulator. Charge disproportionation was identified by the strong multiplet feature developed on the leading edge of Ni $L_3$ edge. our work provides a paradigm for the manipulation of electronic ground states by adjusting the thickness at the nanometer scale, which is helpful for understanding the dimensionality effect on the physical properties of perovskite nickel.

## 4. Experimental Methods

High-quality [(NdNiO$_3$)$_m$/(SrTiO$_3$)$_1$] SLs (m = 1, 2, 3, 4, 5, 7 and 9) were grown in a layer-by-layer mode on (001)-oriented STO substrates using a DCA R450 reactive MBE system. At room temperature, the pseudocubic lattice constants for orthorhombic structure NdNiO$_3$(a = 3.803 Å), which is expected to be subjected to tensile strain on STO substrate (3.905 Å). Atomically flat, TiO$_2$-terminated stepped STO substrates were purchased from SHINKOSHA of Japan, and they were annealed at 300 °C in an ultrahigh vacuum (2×10$^{-9}$ Torr) for half an hour to get rid of the volatile contaminates before growth. High purity element Strontium (Sr), titanium (Ti), neodymium (Nd) and nickel (Ni) were evaporated through thermal Knudsen cells. The flux of each metal source was calibrated prior to film growth at the position of the substrate using a quartz crystal microbalance measurement. During the epitaxy process, the substrate temperature was maintained at 550 °C in a distilled ozone atmosphere of 5×10$^{-6}$ Torr. Here, the distilled ozone was supplied by a DCA Ozone Delivery System. During growth, both the growth rate and surface structure were monitored by in situ reflection high-energy electron diffraction (RHEED).

The structure characterization of the films was carried out in a Bruker-D8 discover x-ray diffractometer (XRD) with Cu κ$_α$ ($\lambda$ = 0.154 nm). In-plane transport measurements were performed by means of the conventional H. C. Montgomery technique in the temperature ranged from 10 K to 400 K using a Quantum Design Physical Property Measurement System (PPMS).[48] Electrodes were made by silver glue or bonding aluminum wires which were directly welded onto the SLs' surfaces. Soft XAS measurements were performed at 02B beamline of Shanghai Synchrotron Radiation Facility (SSRF). The L-edge of Ni and the K-edge of O spectra were collected using the total electron yield mode at the room temperature in an ultrahigh vacuum chamber with a base pressure better than 1×10$^{-9}$ Torr. All results have been calibrated with reference samples which were measured simultaneously.

**Supporting Information**

Supporting Information is available from the Wiley Online Library or from the author.

**Acknowledgements**

X. Lu and J. Liu contributed equally to this work. This work was supported by the National Natural Science Foundation of China (Grant Nos. U2032208 and 11905283). J.S.L. thanks the fund of Science and Technology on Surface Physics and Chemistry Laboratory (6142A02200102). Part of this research used Beamline 02B of the Shanghai Synchrotron Radiation Facility, which is supported by ME$^2$ project under contract No. 11227902 from National Natural Science Foundation of China.

Received: ((will be filled in by the editorial staff))
Revised: ((will be filled in by the editorial staff))
Published online: ((will be filled in by the editorial staff))

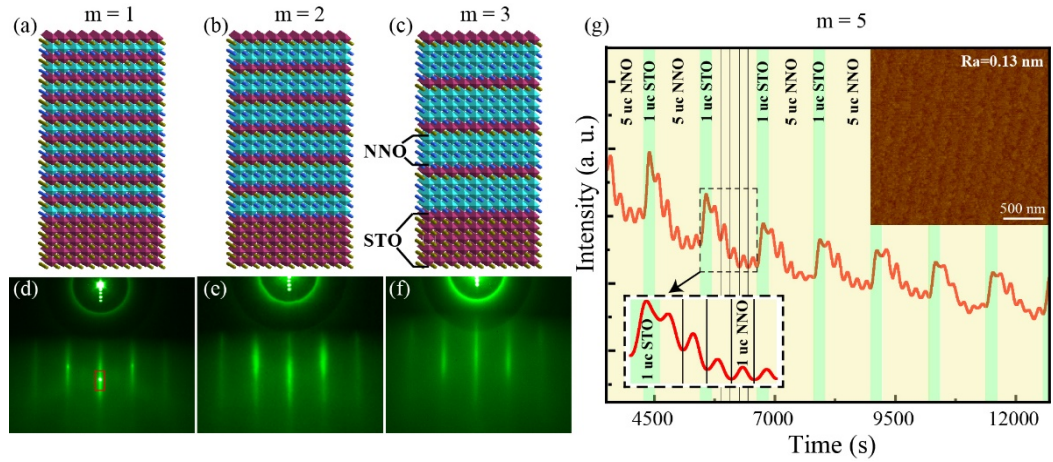

**Figure 1.** (a), (b), (c) Structure schematics of the m = 1, 2, 3 (NdNiO$_3$)$_m$/(SrTiO$_3$)$_1$ superlattices on SrTiO$_3$(001) substrates and their corresponding RHEED patterns (d), (e), (f), respectively. The sharp and clear stripes on RHEED pattern show the in-plane lattice arrangement of epitaxial SLs is very neat. (g) Epitaxial growth time dependence of m = 5 SL RHEED diffraction intensity oscillation curve of [00] diffraction rod, which is marked with a red rectangle window in Figure 1(a). Different colored backgrounds represent the growth of different films which reflects the layer-by-layer growth pattern of SL. The right upper inset is the typical flat morphology measured by Atomic Force Microscope (AFM). (a.u.=arbitrary units).

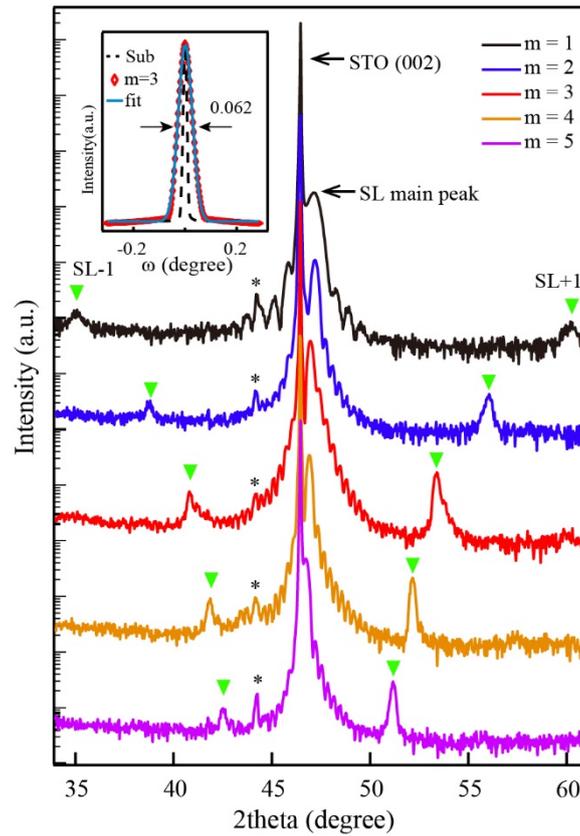

**Figure 2.** (a) The typical film X-ray diffraction ($\theta$-$2\theta$) scans for m = 1, 2, 3, 4, 5 SLs. The black arrows point to the (002) Bragg peak of STO and SLs main peak–near STO (002), respectively. The satellite peaks of SLs are indicated by green triangles and their changes with m can be seen clearly. Black ∗ is the spurious signal of equipment. Inset is the typical rocking curve of m = 3 SL, which is sharp and comparable to that of substrate.

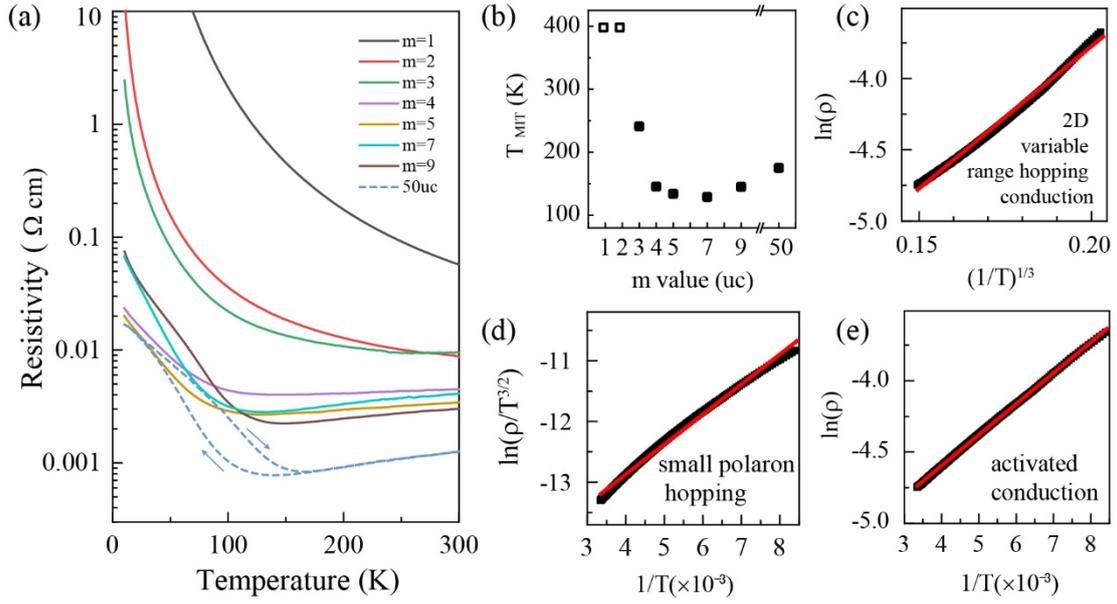

**Figure 3.** (a) Resistivity versus temperature during warming for SLs and 50 uc NNO. (b) $T_{MIT}$ as a function of m value. The hollow box means that the SLs of m = 1, 2 are the insulators within the measurement temperature range. (c), (d), (e) Linear fit (red line) to two-dimensional variable range hopping, small polaron hopping, and activated conduction model for m = 2 SL, respectively.

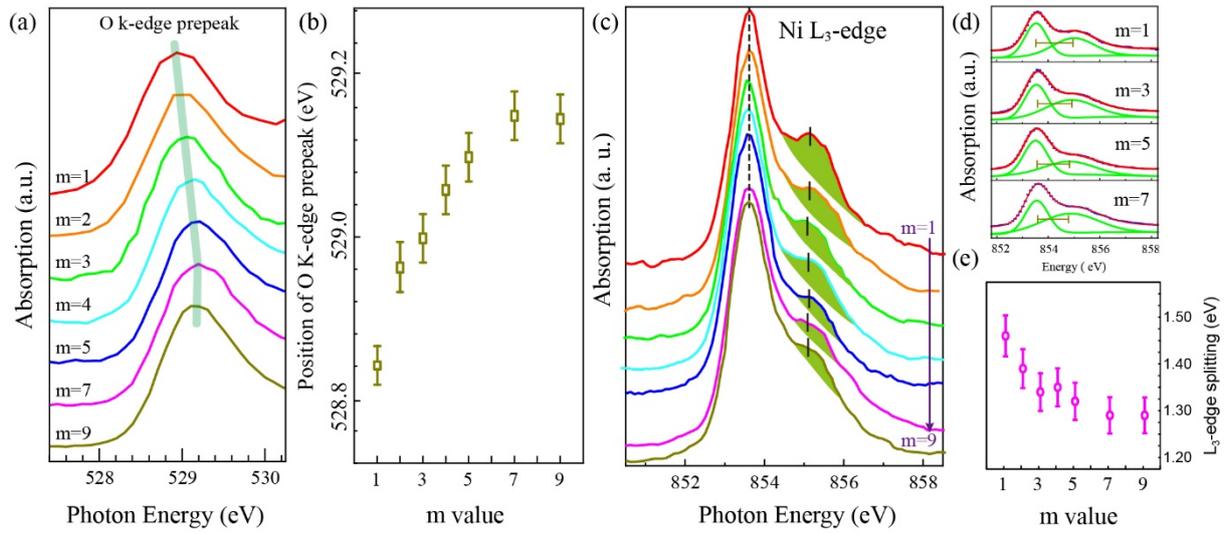

**Figure 4.** (a) Normalized prepeak of O K-edge absorption for $(NNO)_m/(STO)_1$ SLs measured at room temperature. Shaded line is guide for eyes. (b) The position of O K-edge prepeak versus m value of SLs, which determined by the charge-transfer energy between O $2p$ and Ni $3d$ band. Error bars are limited to the photon energy resolution. (c) Normalized Ni $L_3$ edges of the SLs. The shaded green areas highlight the evolution of the multiplet feature. Vertical bars are guide for eyes to view the shift of the peaks. (d) The fitting of Ni $L_3$ spectra with the sum of two peaks for selected m SLs. (e) The splitting energy extracted from fitting of Ni $L_3$ edge for SLs as a function of m value.

Supporting Information

**Dimensionality-controlled evolution of charge-transfer energy in digital nickelates superlattices**

*Xiangle Lu, Jishan Liu*, Nian Zhang, Binping Xie, Shuai Yang, Wanling Liu, Zhicheng Jiang, Zhe Huang, Yichen Yang, Jin Miao, Wei Li, Soohyun Cho, Zhengtai Liu, Zhonghao Liu and Dawei Shen**

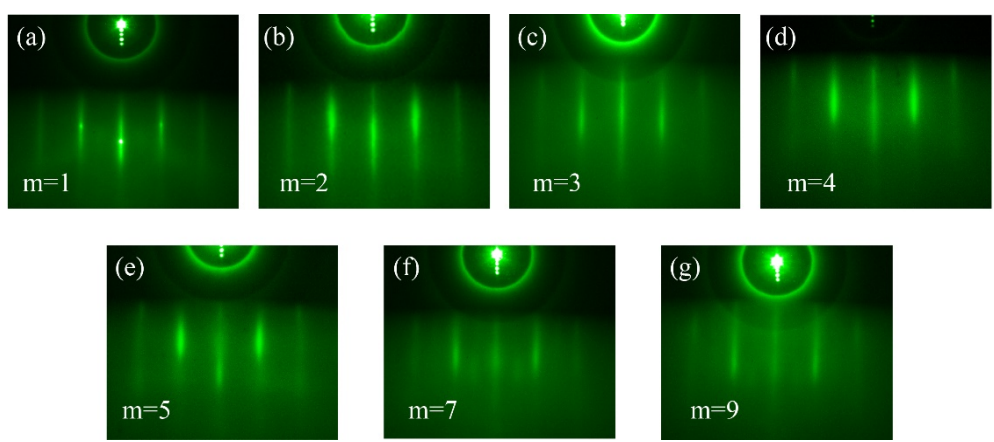

**Figure S1.** (a)(b)(c)(d)(e)(f)(g) RHEED patterns of superlattices.

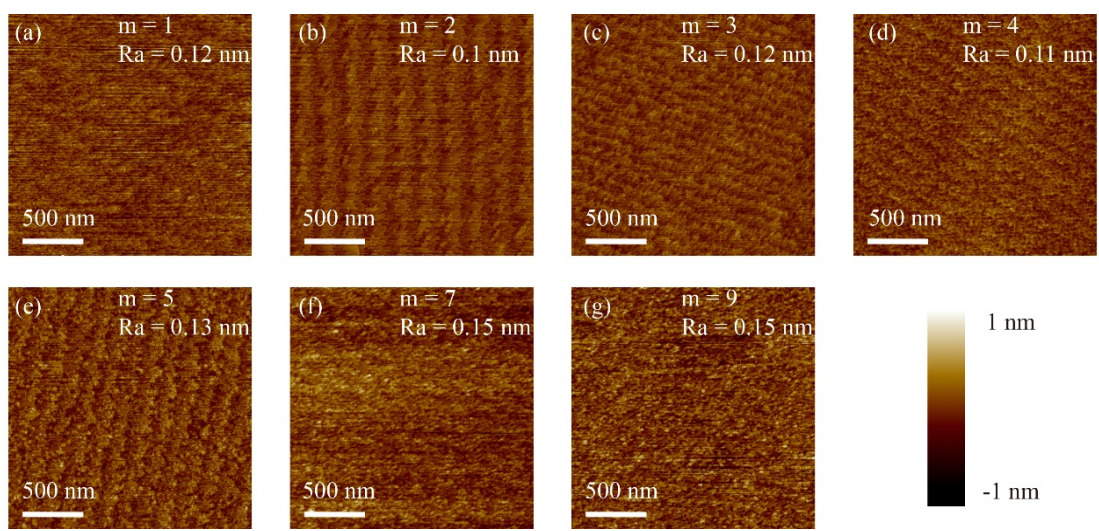

**Figure S2.** (a)(b)(c)(d)(e)(f)(g) AFM image of superlattices. The lower right corner is a color scale. The average surface roughness Ra of the m = 1, 2, 3, 4, 5, 7, 9 superlattices is 0.12 nm, 0.1 nm, 0.12 nm, 0.11 nm, 0.13 nm, 0.15 nm, 0.15 nm, respectively.

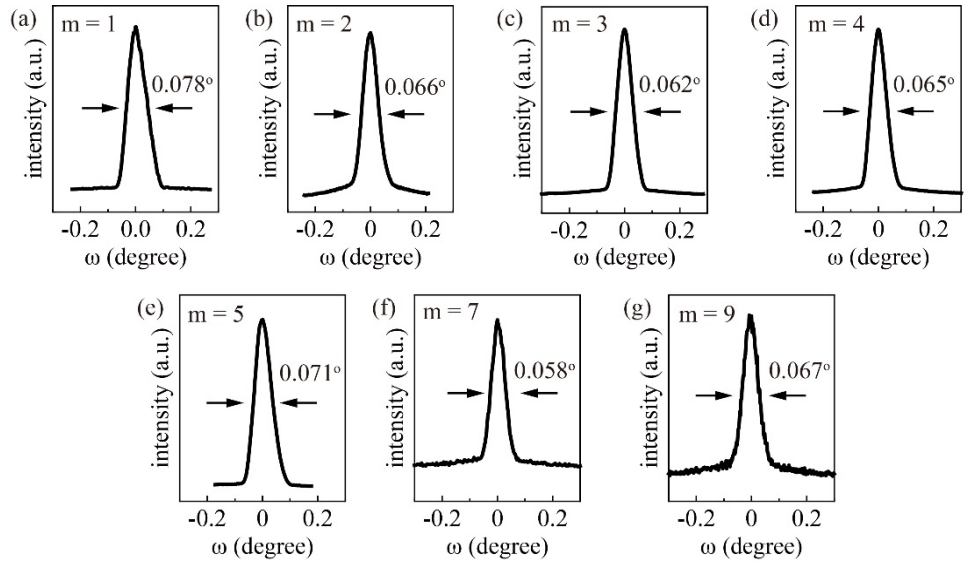

**Figure S3.** (a)(b)(c)(d)(e)(f)(g) Rocking curve of superlattices with XRD. The full width at half maxima (FWHM) of the m = 1, 2, 3, 4, 5, 7, 9 superlattices is 0.078°, 0.066°, 0.065°, 0.065°, 0.071°, 0.058°, 0.067°, through Gaussian fitting, respectively.

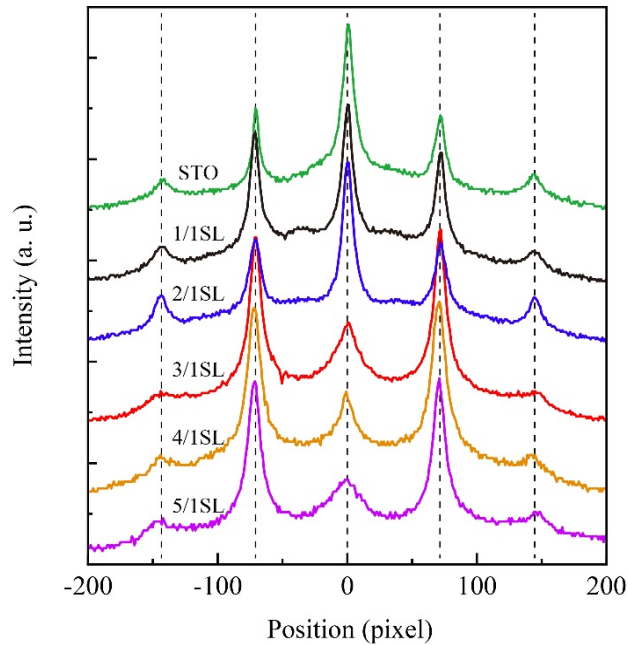

**Figure S4.** Comparison of RHEED intensity integral curves of $SrTiO_3$ and superlattice films. It clearly shows that the RHEED streak spacing of the SLs match well with those of the $SrTiO_3$ substrates, indicating the coherent growth of the films on $SrTiO_3$ substrates.

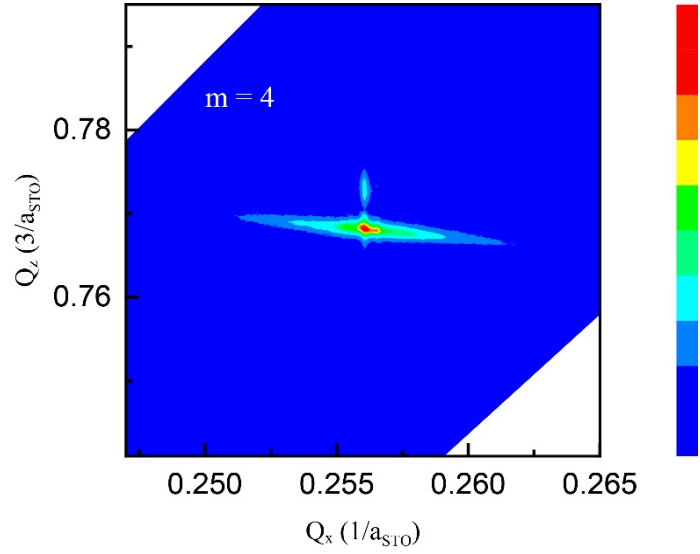

**Figure S5.** The RSM around (103) reflections for m = 4 SL. Along the horizontal axis, the film is in-plane lattice matched to the SrTiO$_3$ substrate, which confirms that the SL film grown on SrTiO$_3$ is fully strained.

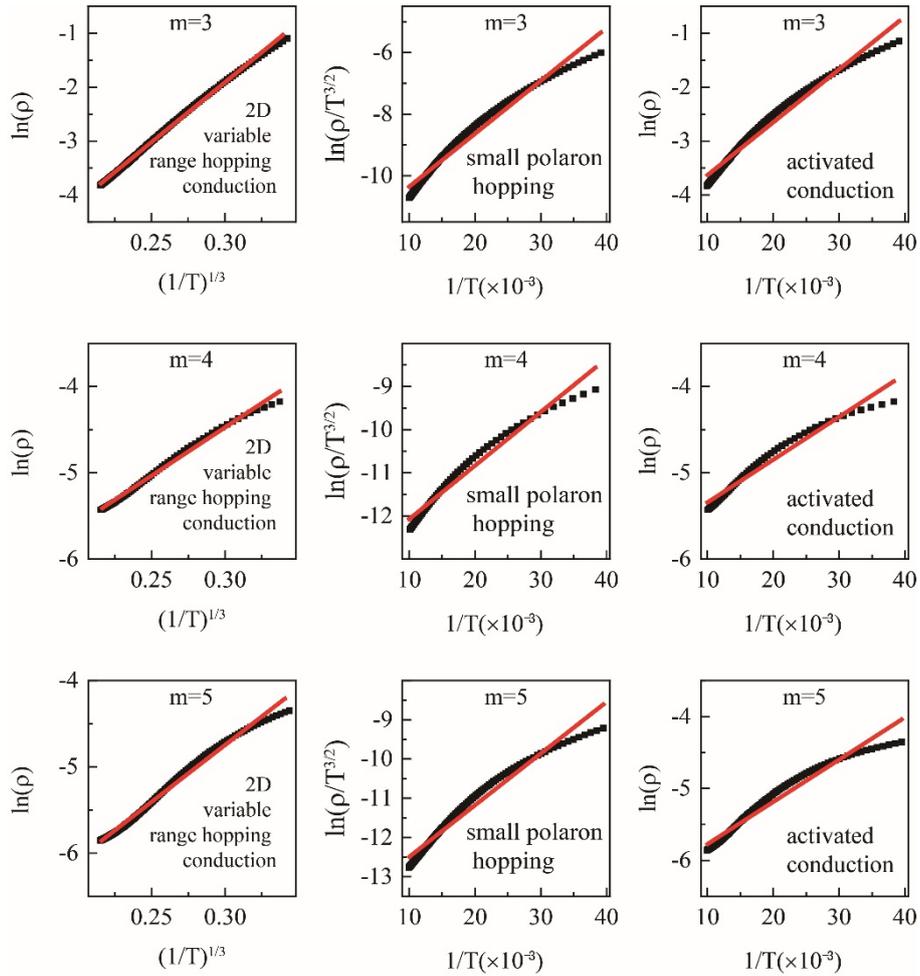

**Figure S6.** Linear fit (red line) to two-dimensional variable range hopping, small polaron hopping, and activated conduction model for m = 3, 4, 5 SLs, respectively. It is difficult to find a suitable single model to fit well due to the complex resistance-temperature curve in the insulating regions for these samples.

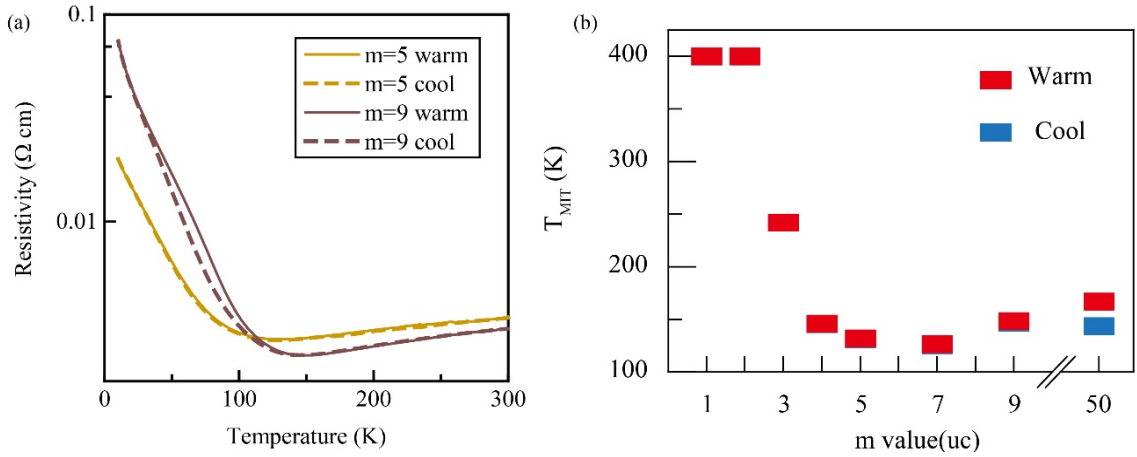

**Figure S7.** (a) Resistivity versus temperature during warming and cooling for m = 5, 9 SLs. (b) The $T_{MIT}$ of the SLs series obtained by derivation $d\rho/dT = 0$ during warming and cooling.

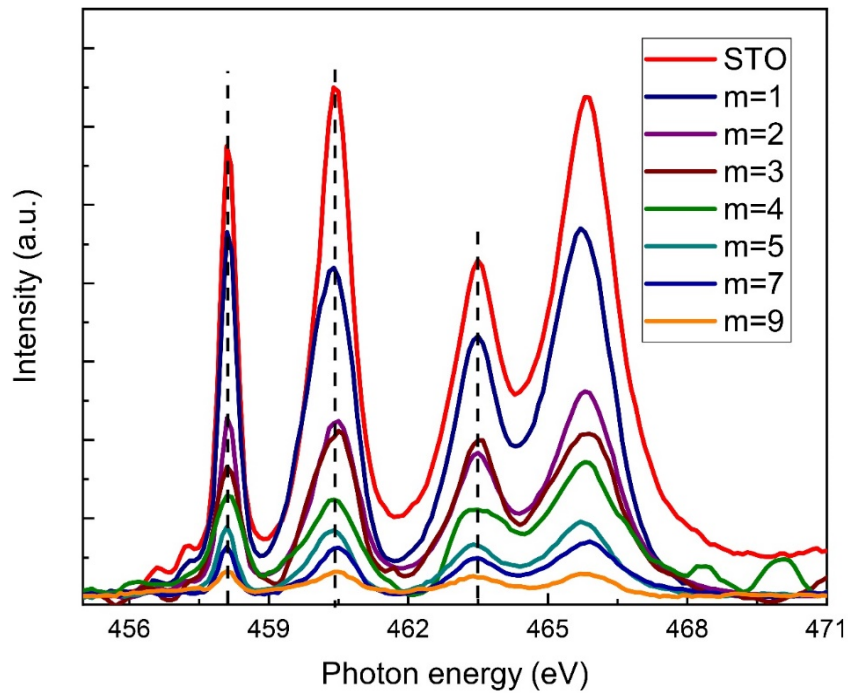

**Figure S8.** (a) XAS at the Ti $L$ edge for the SLs series. No charge transfer happened at the STO/NNO interface for the Ti cation strongly prefers the +4 oxidation state regardless of the thickness of NNO slab.

21